\documentclass[]{spie}
\usepackage{graphicx,cite, fullpage, multirow}
\begin{document}
\title{Modelling the application of integrated photonic spectrographs to astronomy}

\author{R. J. Harris\supit{a} and J. R. Allington-Smith\supit{a}
\skiplinehalf
\supit{a}Centre for Advanced Instrumentation, Phyiscs Department, Durham University, South Rd, Durham, DH1 3LE, UK;}
\authorinfo{Further author information: (Send correspondence to R.J.Harris)\\R.J.H.: E-mail: r.j.harris@durham.ac.uk\\  JRAS.: E-mail: j.r.allington-smith@dur.ac.uk}

\date{}
\maketitle

\newcommand{\mr}{\multirow{2}{*}}

\abstract{One of the well-known problems of producing instruments for Extremely Large Telescopes is that their size (and hence cost) scales rapidly with telescope aperture. To try to break this relation alternative new technologies have been proposed, such as the use of the Integrated Photonic Spectrograph (IPS). Due to their diffraction limited nature the IPS is claimed to defeat the harsh scaling law applying to conventional instruments.
The problem with astronomical applications is that unlike conventional photonics, they are not usually fed by diffraction limited sources. This means in order to retain throughput and spatial information the IPS will require multiple Arrayed Waveguide Gratings (AWGs) and a photonic lantern. We investigate the implications of these extra components on the size of the instrument. We also investigate the potential size advantage of using an IPS as opposed to conventional monolithic optics.
To do this, we have constructed toy models of IPS and conventional image sliced spectrographs to calculate the relative instrument sizes and their requirements in terms of numbers of detector pixels. Using these models we can quantify the relative size/cost advantage for different types of instrument, by varying different parameters e.g. multiplex gain and spectral resolution. This is accompanied by an assessment of the uncertainties in these predictions, which may prove crucial for the planning of future instrumentation for highly-multiplexed spectroscopy.}

\section{Introduction}

Spectroscopy is the most useful technique in modern Astronomy, addressing problems ranging from fundamental cosmology to exoplanets. In its simplest form a dispersive spectrograph contains four components. A slit to isolate the area to be dispersed, a collimator, a grating or prism to disperse the light and a camera and detector to record the intensity at each wavelength. In order to retain  throughput the slitwidth of a spectrograph is usually matched to the FWHM of the seeing. If the system is not diffraction limited then the collimated beam must increase in size in order to maintain the same resolution \cite{Lee2000}. This means that the disperser must also grow in size. This leads physical problems such as stresses and flexure in the materials, along with the difficulties inherent in building these large monolithic structures. In order to use the same design principles as existing instruments more exotic materials and construction processes are needed, which drives the costs of building the instruments up. This leads to the cost of instruments scaling with the square of the telescope aperture or more \cite{Bland-Hawthorn2006}. \newline
Many ways to break this relationship have been considering including image slicing \cite{Allington-Smith2009} and adopting technologies developed by the photonics industry \cite{Bland-Hawthorn2006}. The solution considered in this paper makes use of the Arrayed Waveguide Gratings (AWGs) as the dispersive element. These form part of the Integrated Photonic Spectrograph (IPS) concept \cite{Bland-Hawthorn2010}. \newline
The IPS takes light from an input fibre which is usually matched to the seeing limit (as with conventional fibre fed instruments) and so supports many modes. As the input to the AWG must be single moded the light has to be split into individual single moded fibres by a photonic lantern \cite{Noordegraaf2010}. These are then fed into the AWGs which disperse the light into individual spectra. \cite{Bland-Hawthorn2010}. \newline
While the principles have been demonstrated \cite{Cvetojevic2009, Cvetojevic2012}, so far these have only used single or few modes from a single input, so have suffered from loss of throughput and coverage of field.

In this paper we shall look at whether photonic spectroscopy can compete with conventional instruments in terms of size. We will construct a toy model of conventional instruments and photonic ones in order to evaluate the relative volumes produced by these instruments.

\section{Instrument size scale relationships}

It has been stated in Astrophotonics papers  that using AWGs breaks the telescope diameter, spectral resolving power relation\cite{Bland-Hawthorn2010}. The relation, 

\begin{equation}
	R =  \frac{m \rho \lambda W}{\chi D_{T}}  = \frac{2 \tan \gamma D_{col}}{\chi D_{T}} 
	\label{eqn:res}
\end{equation}

holds for a long slit spectrograph using a diffraction grating as the dispersive element. Here \textit{R} is the resolution of the instrument, m is the diffraction order, $\lambda$ is the wavelength, \textit{W} is the length of intersection between the grating and collimated beam, $\chi$ is the angular slitwidth, $D_{T}$ is the diameter of the telescope, $\gamma$ is the blaze angle and $D_{col}$ the diameter of the collimated beam.  It shows that for a given resolution, angular slitwidth, and blaze angle, that as the diameter of telescope increases the diameter of the collimated beam must increase, leading to a bigger instrument. Unlike a conventional spectrograph the AWG must be diffraction limited ($\lambda = \chi D_{T}$) due to its single mode nature, so the spectral resolving power can be shown to be of the form 

\begin{equation}
	R  = \frac{mN}{C}.
	\label{eqn:awg_res}
\end{equation}

where \textit{N} is the number of waveguides, \textit{m} is the diffraction order and \textit{C} errors from manufacturing \cite{Lawrence2010a}. It must be noted that this is for a single AWG. When telescope is not at the diffraction limit multiple AWGs (and hence modes) would be needed to conserve throughput. 

The number of spatial modes, in a stepped index fibre can be approximated as

\begin{equation}
	M = \frac{V_{fibre}^{2}}{4} .
\end{equation}

The associated V parameter being given by 

\begin{equation}
	V_{fibre} = \frac{ \pi s \Theta}{\lambda} .
\end{equation}
Where \textit{s} is the diameter of the fibre, and $\Theta$ the numerical aperture at which the fibre is operated \cite{cheo}. This must be less than the maximum numerical aperture of the fibre. This can be modified so that the variables so they are same as Eq. (\ref{eqn:res}) by taking

\begin{eqnarray}
	\Theta &\approx& \frac{1}{2 F_{T}} \\
	s &=& \chi f_{T} = \chi F_{T} D_{T}  \\
\end{eqnarray} 

\noindent with $F_{T}$ being the focal ratio, $f_{T}$ the focal length of the telescope and s the diameter of the fibre (equivalent to the slitwidth in Eq. (\ref{eqn:res})), yielding

\begin{equation}
	M =  \left( \frac{\pi \chi D_{T}}{4 \lambda} \right)^{2}.
	\label{eqn:modes}
\end{equation}

Therefore it can be seen that for each sampling element the number of modes increases as the square of the telescope diameter in a similar way to the number of slices at the diffraction limit $(\chi D_{T} / 1.22 \lambda )^{2}$. This means unless we will need to utilise more AWGs in order to maintain throughput, therefore the size of the IPS will still scale with telescope diameter. \newline

%

\section{Modelling}

We model both the standard instrument and its AWG equivalent. For our standard instrument we use a double pass echelle model similar to that developed in Ref. \citenum{Allington-Smith2009}. We choose a variable focal length for the camera (and hence telescope) to keep the model simple and compact. For our AWG model we choose a reflective design similar to those described in Refs.  \citenum{Peralta2003} and \citenum{DePeralta2004}. \newline
The major difference between the models is the way they are scaled. All other parameters are based upon existing instruments for both models. As such extra components such as the optics, detector and housing are accounted for in the echelle model simply by using scaling factors. The instrument itself will have been optimised for size and structural integrity so it will represent the optimal instrument design. \newline
By comparison there are no full scale photonic spectrographs currently in existence. The current generation are prototypes, only sampling a very small area of the sky and not making use of enough AWGs to allow high throughput. Because of this we use a bottoms up approach when scaling our AWG model. We use an existing AWG obtained from Gemfire Livingstone and use its known size to scale our model. This does however mean the AWG model does not include the extra equipment that would be required for a full instrument, such as detectors and the housing of the instrument. We also assume all the AWGs in the final device can be stacked on top of each other, which in practise may not be possible.

\subsection{The echelle Model}

\begin{figure}
	\begin{center}
		\includegraphics[width = 0.9\textwidth]{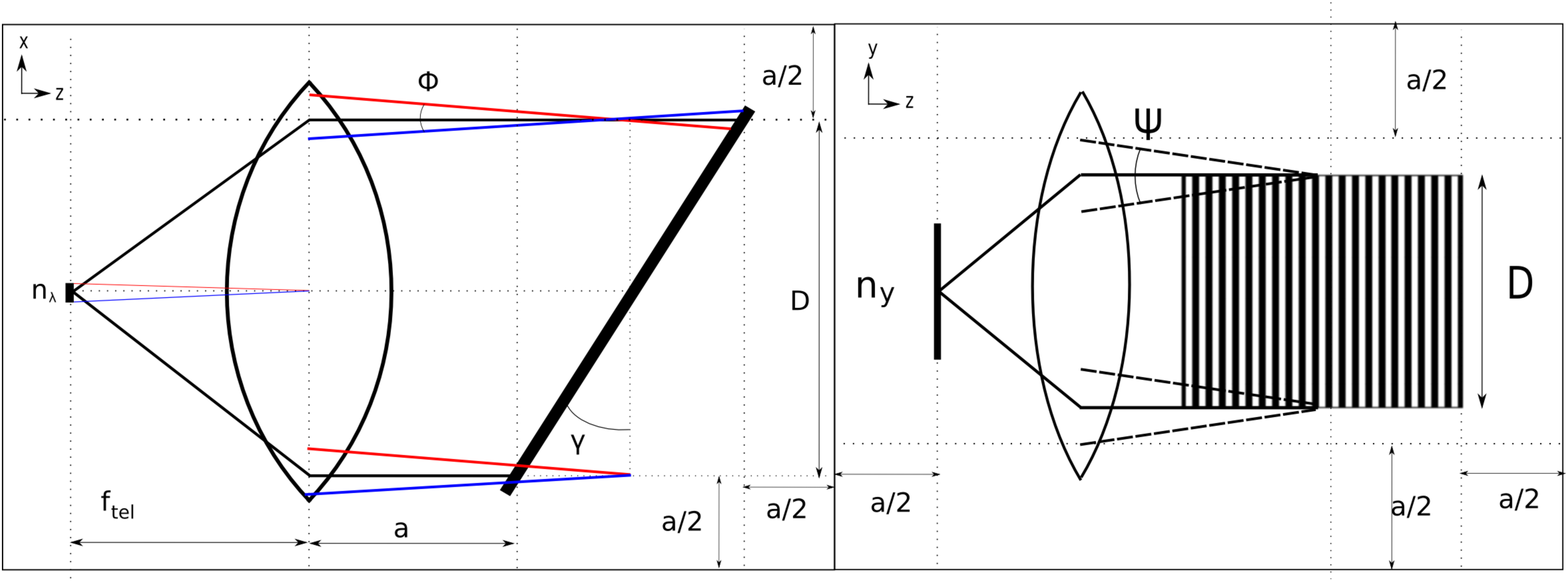}
		\caption{The echelle model. The left part of the image shows the spectra dispersion of the echelle model, and the right part shows the spatial extent of the device. This is identical to the one described in Ref. \citenum{Allington-Smith2009a}}
		\label{fig:echelle}
	\end{center}
\end{figure}

The echelle model is adapted from that shown in Ref. \citenum{Allington-Smith2009a}. Here we will be calculating the model using the initial input parameters of $\chi$, R,  $\lambda_{c}$,  $\Delta \lambda_{FSR}$,  $N_{y}$,  $\rho$, where $\Delta \lambda_{FSR}$ is the free spectral range (FSR). 
As stated in Ref. \citenum{Allington-Smith2009a} modelling a spectrograph for monochromatic light is simple, but for a large wavelength range we need to add in extra parameters accounting for dispersion. For our model here we use $\pm \Phi/2$ to represent the extra beamspread due to the wavelength dispersion and $\pm \Psi/2$ to represent the dispersion due to conservation of etendue.
Our toy model is based on an echelle spectrograph with double pass construction (see Fig. \ref{fig:echelle}). For an unsliced spectrograph its dimensions are estimated as

\begin{eqnarray}
	L_{x} &=& \left( D_{col} + \Phi \left( a_{ech} + D_{col}\tan \gamma \right) + a_{ech} \right)b_{ech} \\
	L_{y} &=& \left( D_{col} + \Psi \left( a_{ech} + D_{col}\tan \gamma \right) + a_{ech} \right)b_{ech} \\
	L_{z} &=& \left( D_{col}F_{c} + D_{col}\tan \gamma + 2a_{ech} \right) b_{ech}.  
\end{eqnarray}
Where $D_{col}$ is the diameter of the collimated beam, $\gamma$ the blaze angle and $F_{c}$ the focal ratio of the collimator and camera. We include $a_{ech}$ and $b_{ech}$ as an oversizing parameter in order to allow the model to be fitted to existing instruments, with $a_{ech}$ as the additive term and $b_{ech}$ the multiplicative.
We consider Littrow configuration at blaze, so we can re-arrange Eq. (\ref{eqn:res}) to give the diameter of the collimated beam

\begin{equation}
	D_{col} = \frac{R \chi D_{t}}{2 \tan \gamma}. 
\end{equation}
The blaze angle in Littrow configuration is $\sin \gamma = \frac{\rho \lambda_{c} m}{2}$, and the FSR of a system is $\Delta \lambda_{FSR} = \frac{\lambda_{c}}{m}$. Thus

\begin{equation}
	T = \tan \gamma = \tan \left( \sin^{-1} \left( \frac{\rho \lambda_{c}^{2}}{2 \Delta \lambda_{FSR}} \right) \right).
\end{equation}

To work out the spatial beamspread we use conservation of etendue, yielding
\begin{equation}
	\Psi =  N_{y}\chi \frac{D_{T}}{D_{col}}.
\end{equation}
We now work out $\Phi$, using the FSR and $\lambda_{c}$. In order to calculate the total angle occupied on the detector by the spectrum we use the grating equation $\sin(\theta_{i}) + \sin(\theta_{o}) = m \rho \lambda$. Setting $\theta_{i}$ = $\gamma$ gives

\begin{equation}
	\Phi = \theta(\lambda_{max}) - \theta(\lambda_{min}).
\end{equation}
Where $\lambda_{max}$ and $\lambda_{min}$ are the maximum and minimum wavelengths that the system operates at (= $\lambda_{c}$ $\pm$ $\Delta_{\lambda}$/2). In order to find the minimum focal length for our system we need to make sure we can resolve each point in our spectra. We know that  $\chi F_{c} D_{T} = n_{0} d_{p}$, where $n_{0}$ is the oversampling of the detector (we use 2.5 allowing for a non-gaussian shape in the point spread function (PSF)) and $d_{p}$ is the size of each pixel (our default is 13.5$\mu$m). From this we can calculate the minimum focal ratio of the camera $F_{c}$.

With all this in place we can rewrite the equations for the scale lengths of the instrument as

\begin{eqnarray}
	L_{x} \left(\chi \right) &=& \left( \frac{R \chi D_{T}}{2T} + \Phi \left(a_{ech} + \frac{R \chi D_{T}}{2} \right) + a_{ech} \right) b_{ech}  \\
	L_{y} \left(\chi \right)  &=&  \left( \frac{R \chi D_{T}}{2T} + \Psi \left(a_{ech} + \frac{R \chi D_{T}}{2} \right) + a_{ech} \right) b_{ech}   \\
	L_{z} \left(\chi \right)  &=&  \left( \frac{R \chi D_{T} }{2} \left( \frac{F_{c}}{T} + 1 \right) + 2a_{ech} \right) b_{ech} .
\end{eqnarray}

Finally to correctly sample the output we require one sampling point in the spatial direction (as we have fixed our slit the the FWHM already) and $n_{0}$ samples in our spectral direction, yielding

\begin{equation}
	N_{pixels} = n_{0} N_{y} N_{\lambda} =  \frac{\Delta \lambda_{FSR} n_{0}}{\Delta \lambda} = \frac{\Delta \lambda_{FSR} R n_{0}}{\lambda_{min}}.
\end{equation}	
 
\subsection{The AWG Model}

\begin{figure}
	\begin{center}
		\includegraphics[width = 0.9\textwidth]{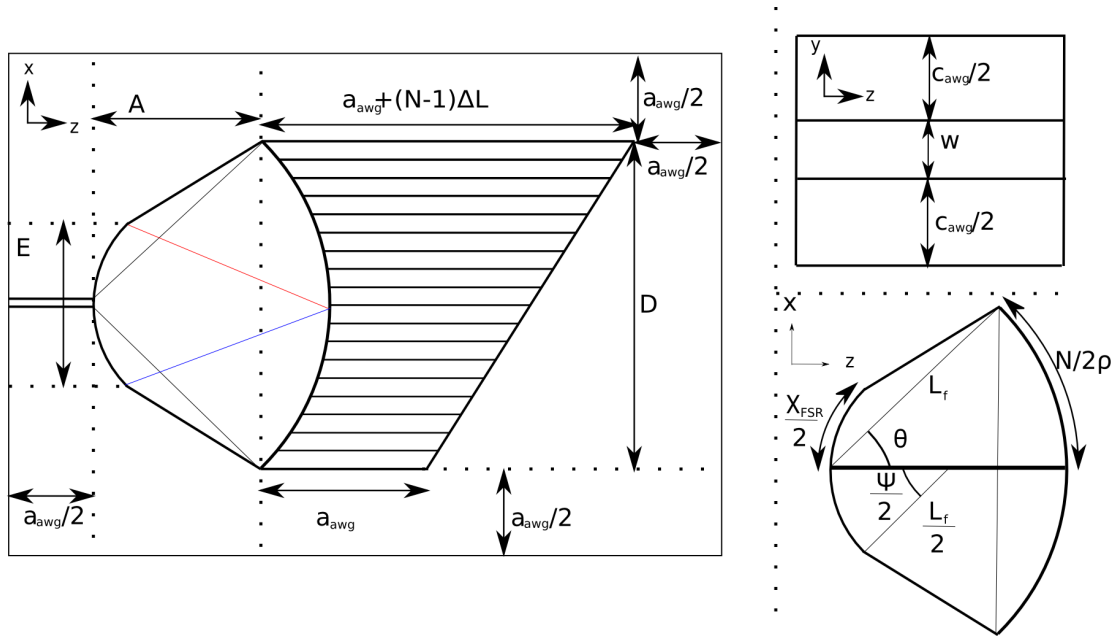}	
		\caption{The AWG model, Left is the x-y view of the AWG, the top right shows the side view of the AWG, with the waveguide in the center. Bottom right is an enlargement of the FPR.}
		\label{awg_model}
	\end{center}
\end{figure}

AWGs are usually purpose built and their design is hugely varied, especially when it comes to arrangement allowing the path difference between waveguides. To keep the toy model simple we have chosen a reflective AWG with Rowlands Circle arrangement for the Free Propagation Region (FPR) \cite{Peralta2003, DePeralta2004}. We have removed the bend at the reflective end of the waveguide array for simplicity (see Fig. \ref{awg_model}). Because of this it looks almost identical to a double pass echelle spectrograph model described above.

Using the definitions in Fig. \ref{awg_model} we arrive at the following equations for the size of the model AWG

\begin{eqnarray}
	L_{x} &=& \left( max(D,E) + a_{awg}   \right)b_{awg} \\
	L_{y} &=&  (c_{awg} +  w)b_{awg} \\
	L_{z} &= &  \left( a_{awg} + A + (N_{wg}-1)\Delta L  \right)b_{awg}. 
\end{eqnarray}

Here $N_{wg}$ is the number of waveguides, \textit{A} is the \textit{x} length of the FPR,  $\Delta L$ the length difference between adjacent waveguides to achieve the required order for a given $\lambda$, \textit{D} is the length containing the waveguides (analogous to the illuminated length of the grating in the echelle model), E the \textit{x} length of the detecting surface, \textit{w} is the waveguide diameter. The oversizing parameters are $a_{awg}$, $b_{awg}$ and $c_{awg}$.

First we calculate the appropriate dispersion order we use the re-arranged FSR equation for an AWG

\begin{equation}
	m =  \frac{\lambda_{min}}{\Delta \lambda_{FSR}}.\\
	\label{eqn:fsr}
\end{equation}

Here \textit{m} is the diffraction order and $n_{c}$ is the refractive index of the waveguides. This can be combined with Eq. (\ref{eqn:awg_res}) and $D = N / \rho$ yielding

\begin{equation}
	D = \frac{C R}{m \rho}. 
\end{equation}

We can make use of the equation for free spatial range in an arrayed waveguide grating ($X_{FSR} = \lambda_{min} L_{F}  \rho/ n_{s} $), where $n_{s}$ is the refractive index of the FPR and $L_{f}$ the length of the free space propagation region. Combining with geometrical arguments gives

\begin{equation}
	E = L_{f} \sin \left( \frac{\lambda_{min} \rho}{n_{s}} \right).
\end{equation}

In order to calculate $\Delta L$ we make use of the equation for calculating the central wavelength of the AWG 

\begin{equation}
	\Delta L = \frac{\lambda_{c} m}{n_{c}}
	\label{eqn:delta_l}
\end{equation}

 $\lambda_{c}$ the central operating wavelength. \textit{A} can be calculated using geometry and is

\begin{equation}
	A = L_{f} \cos( \theta) = L_{f} \cos \left( \frac{N}{2\rho L_{f}} \right).
	\label{A_eqn}
\end{equation}

Where \textit{N} is calculated using Eq. (\ref{eqn:awg_res}). In order to calculate $L_{f}$ we make use of the fact the imaging requires the number of detector pixels to be able to adequately sample the PSF at the resolution required (equivalent to sampling of the echelle model in Ref. \citenum{Allington-Smith2009}). To do this we take the dispersion relation

\begin{equation}
	\left( \frac{\delta \lambda}{\delta x} \right) = \frac{n_{s}}{L_{f}m \rho}.
	\label{eqn:dl_dx}
\end{equation}
This can be combined with Eq. (\ref{eqn:fsr}) to remove \textit{m}. We take the equation for resolution $\delta \lambda$ = $\lambda_{min} / R$ and $\delta_{x} = n_{0} d_{p}$. , We use $\lambda_{min}$ in order to get the maximum $L_{f}$ This allows Eq. (\ref{eqn:dl_dx}) to  become

\begin{equation}
	L_{f} \geq \frac{n_{s} n_{0} d_{p} R \Delta \lambda_{FSR} }{\rho \lambda_{min}^{2}}.
	\label{eqn:Lf}
\end{equation}

Finally we can calculate the number of pixels we need for the required resolution

\begin{equation}
	N_{pixels} = \frac{X_{FSR}}{n_{min} d_{p}} = \frac{\lambda_{c} L_{f} \rho}{ n_{s} d_{p}}.
\end{equation}

\section {Results}

We set the input parameters in Table \ref{tab:instruments} to  represent GNIRS \cite{2006MNRAS.371..380A}, using those listed in Ref. \citenum{Allington-Smith2009} as a guide. We have selected GNIRS as it is representative of an existing facility-class instrument. We then vary resolution, field of view (FOV) of the instrument, telescope diameter and seeing. \newline

\begin{figure}[h]
	\hfill
	\begin{minipage}[t]{.45\textwidth}
		\begin{center}
			\includegraphics[width = 1.0\textwidth]{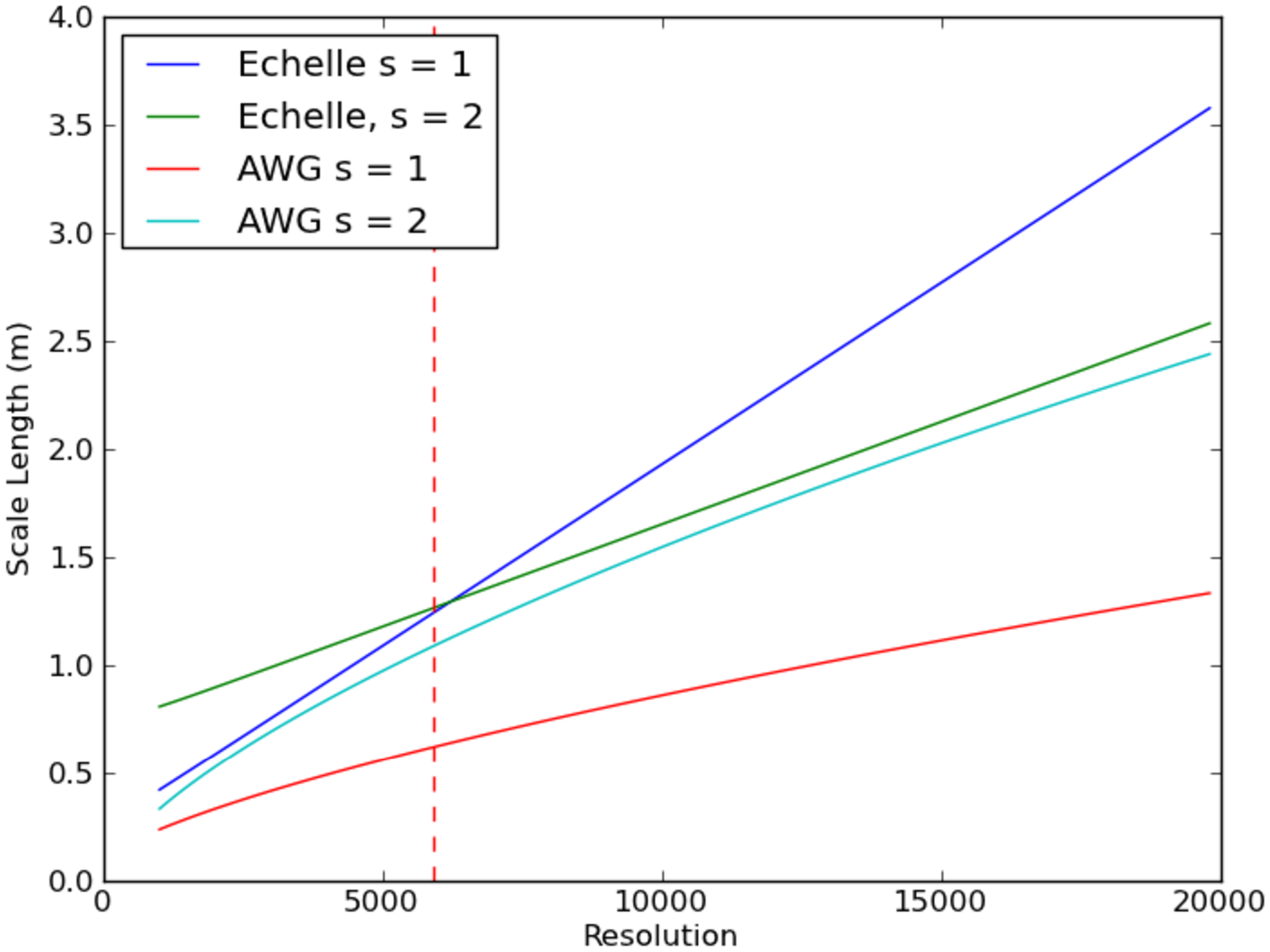}
			\caption{The resulting scale lengths due to varying the resolution of the GNIRS models with the different scaling parameters. This shows the AWG instrument will be smaller over all resolutions.The dashed line represents GNIRS.}
			\label{fig:resolution}
		\end{center}
	\end{minipage}
	\hfill
	\begin{minipage}[t]{.45\textwidth}
		\begin{center}
			\includegraphics[width = 1.0\textwidth]{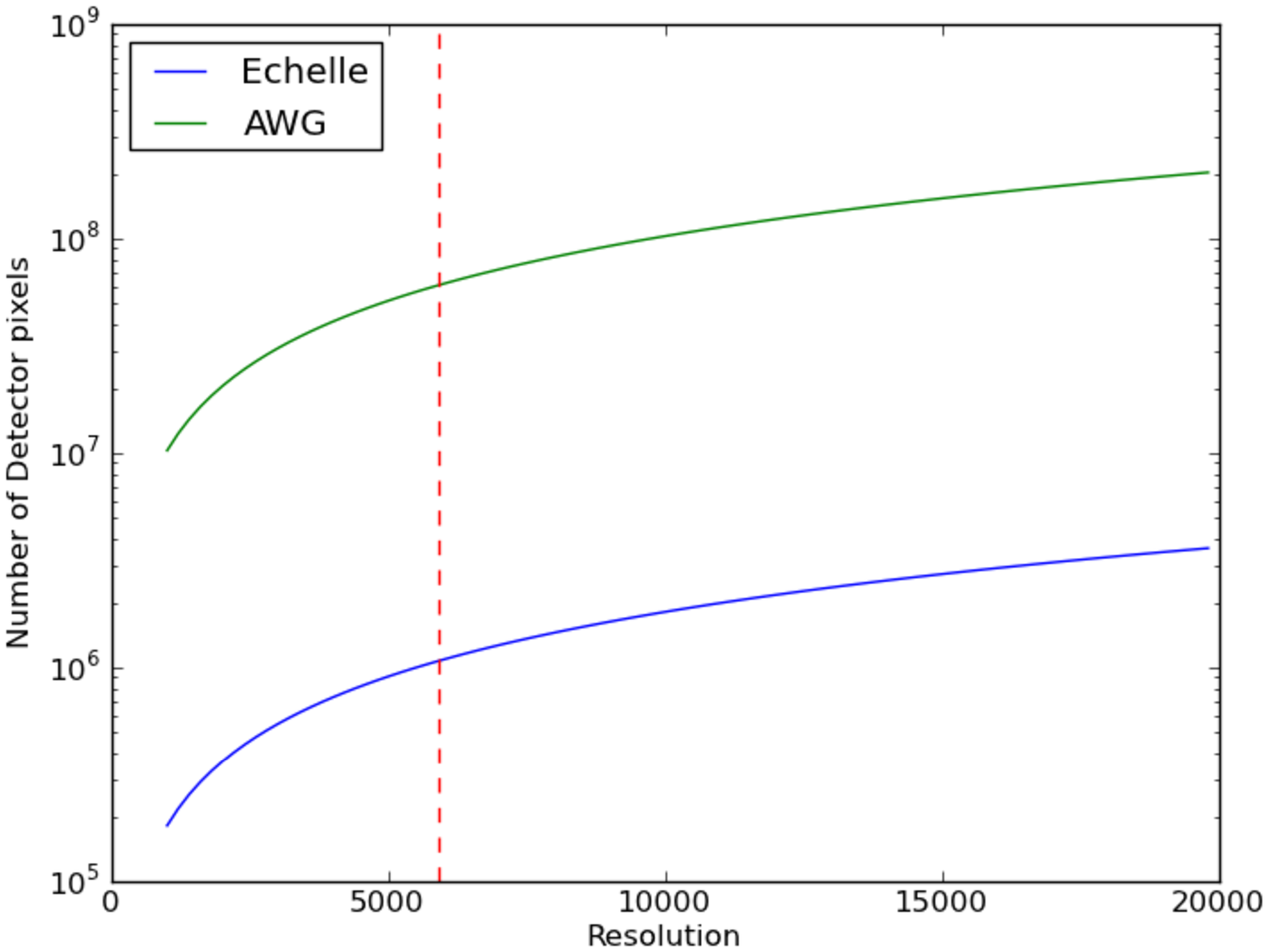}
			\caption{The resulting pixel numbers due to varying the resolution of the GNIRS models with the different scaling parameters. This shows the AWG instrument requires orders of magnitude more detector pixels.The dashed line represents GNIRS.}
			\label{fig:res_pixels}
		\end{center}
	\end{minipage}
	\hfill
\end{figure}

\begin{figure}[h]
	\hfill
	\begin{minipage}[t]{.45\textwidth}
		\begin{center}
			\includegraphics[width = 1.0\textwidth]{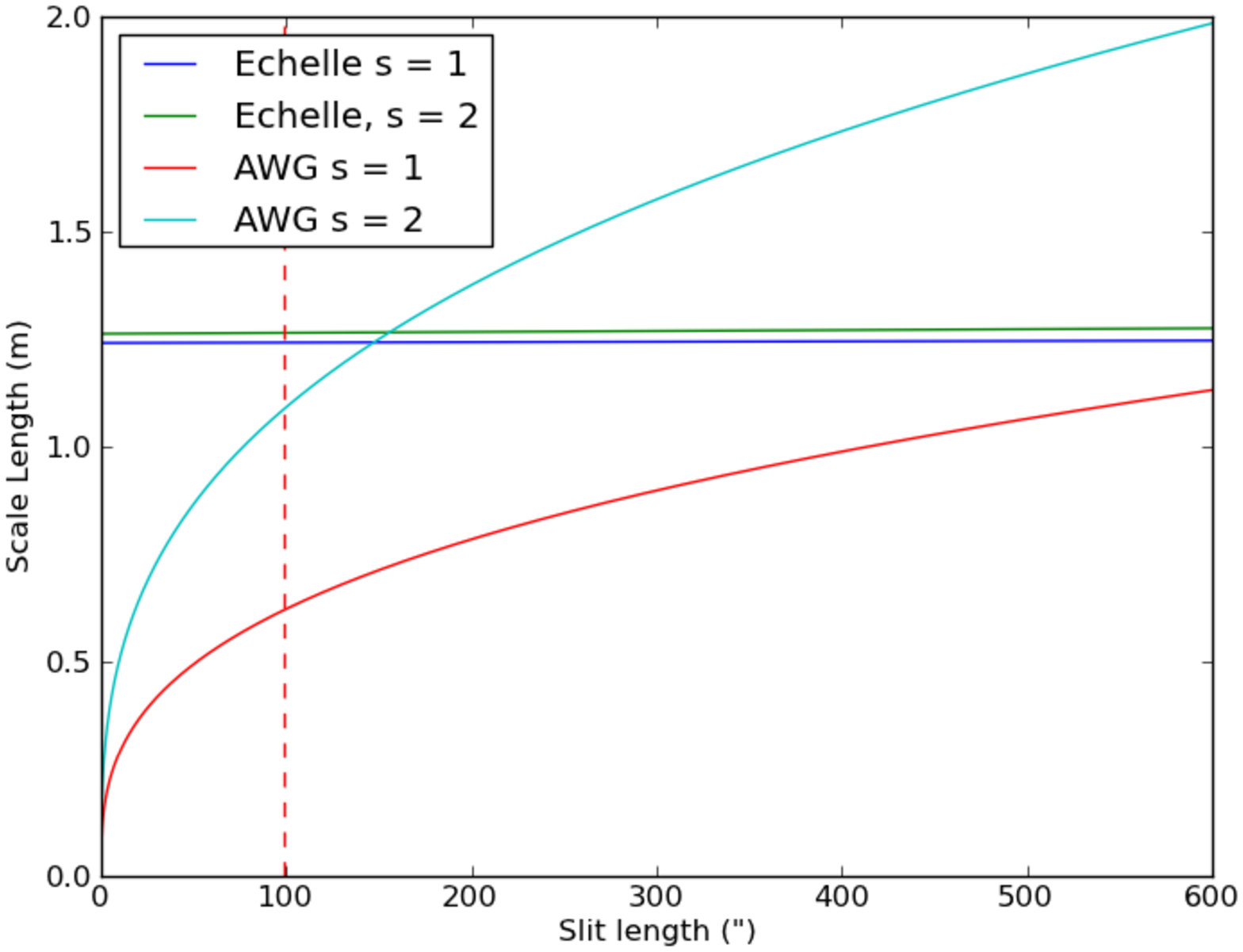}
			\caption{The resulting scale lengths due to varying the slit length of the model instruments. From this we can see that increasing the slit length will have a far greater effect on the AWG model than the conventional instruments. The dashed line represents GNIRS.}
			\label{fig:fov}
		\end{center}
	\end{minipage}
	\hfill
	\begin{minipage}[t]{.45\textwidth}
		\begin{center}
			\includegraphics[width = 1.0\textwidth]{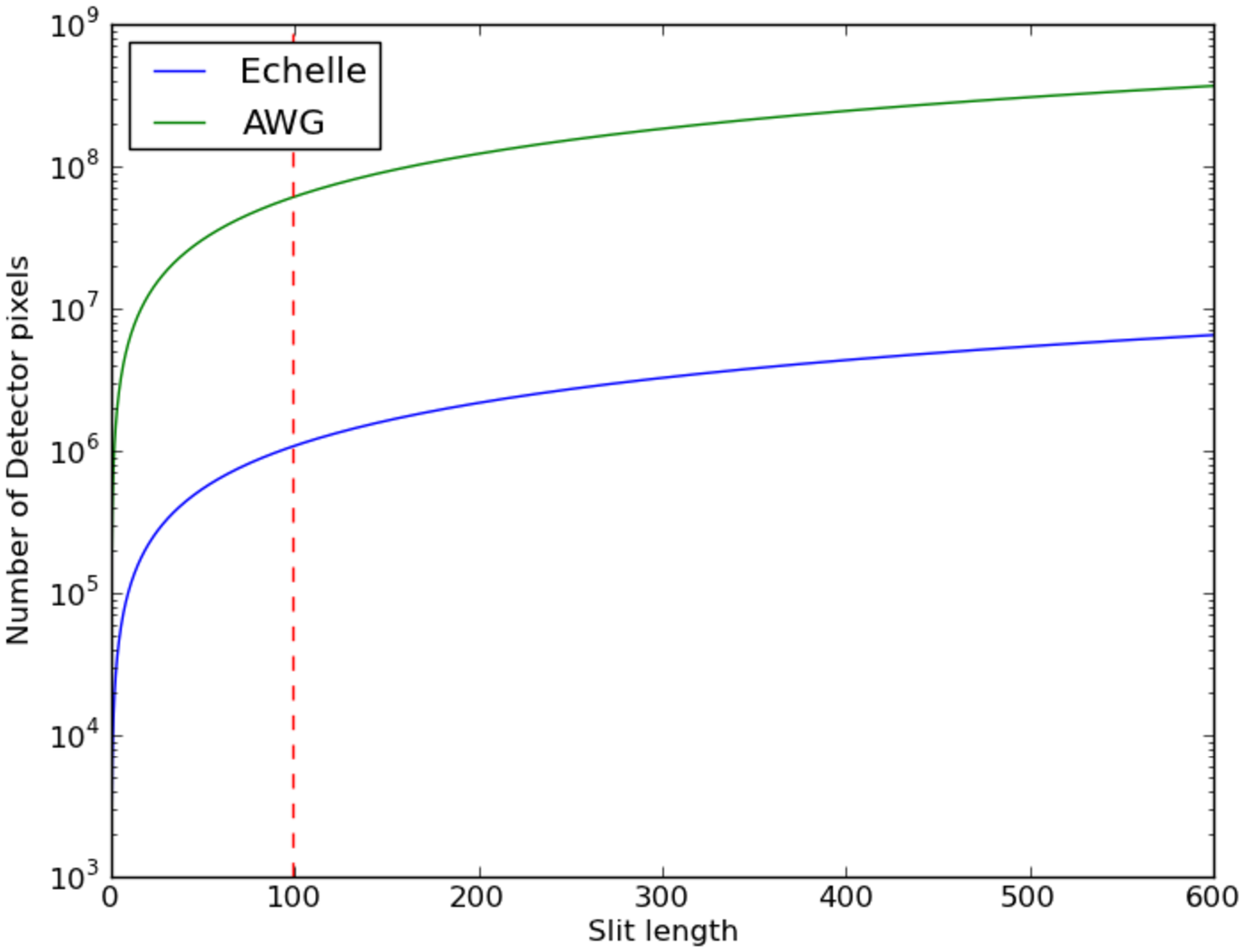}
			\caption{The resulting pixel numbers due to varying the slit length of our model instruments. From this we can see that increasing the length will have a far greater effect on the AWG model than the conventional instruments. The dashed line represents GNIRS.}
			\label{fig:fov_pixels}
		\end{center}
	\end{minipage}
	\hfill
\end{figure}

\begin{figure}[h]
	\hfill
	\begin{minipage}[t]{.45\textwidth}
		\begin{center}
			\includegraphics[width = 1.0\textwidth]{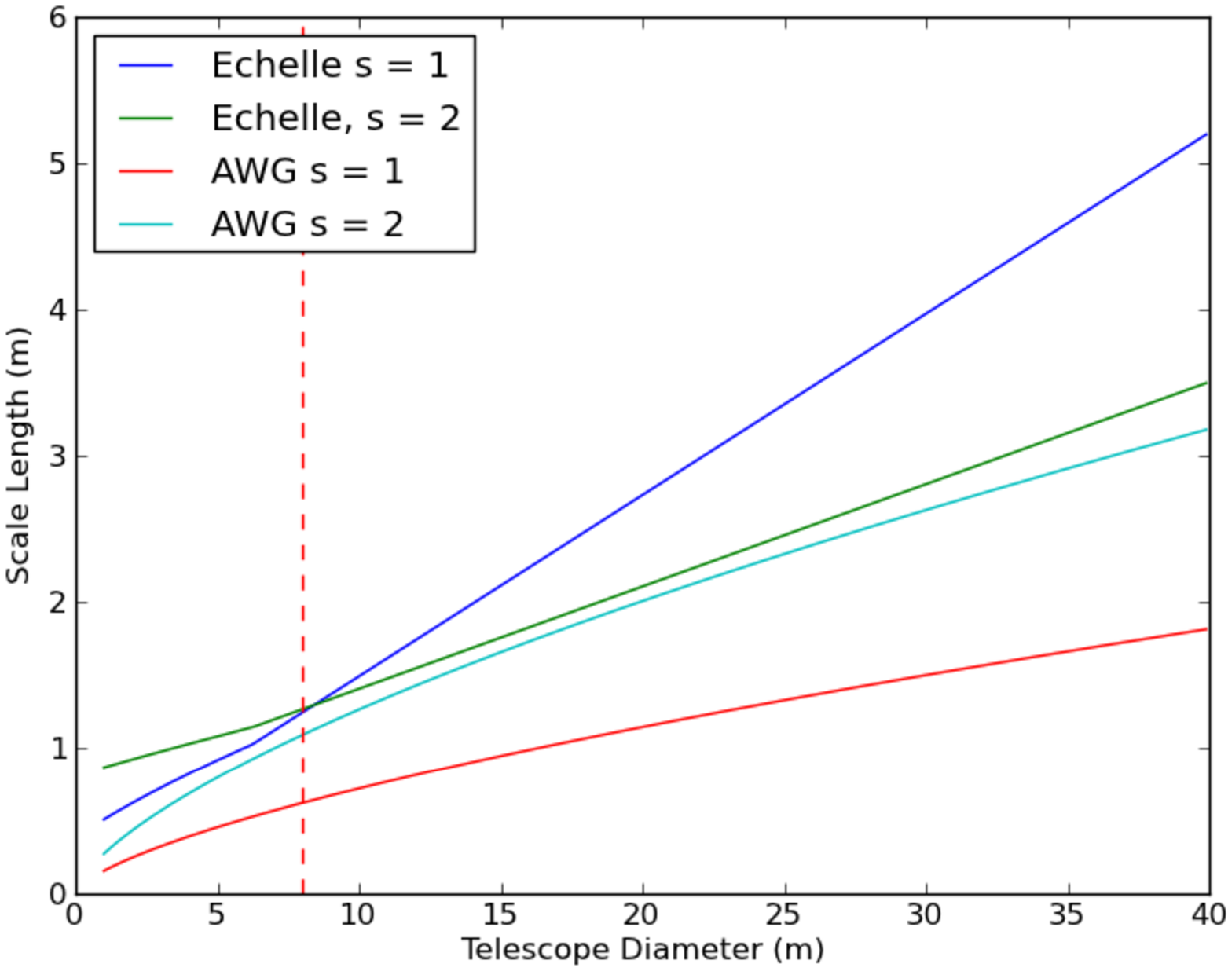}
			\caption{The resulting scale lengths due to varying the diameter of the telescope feeding our model instrument. The difference in instrument sizes is not greatly influenced by telescope diameter. The dashed line represents GNIRS.}
			\label{fig:d_tel}
		\end{center}
	\end{minipage}
	\hfill
	\begin{minipage}[t]{.45\textwidth}
		\begin{center}
			\includegraphics[width = 1.0\textwidth]{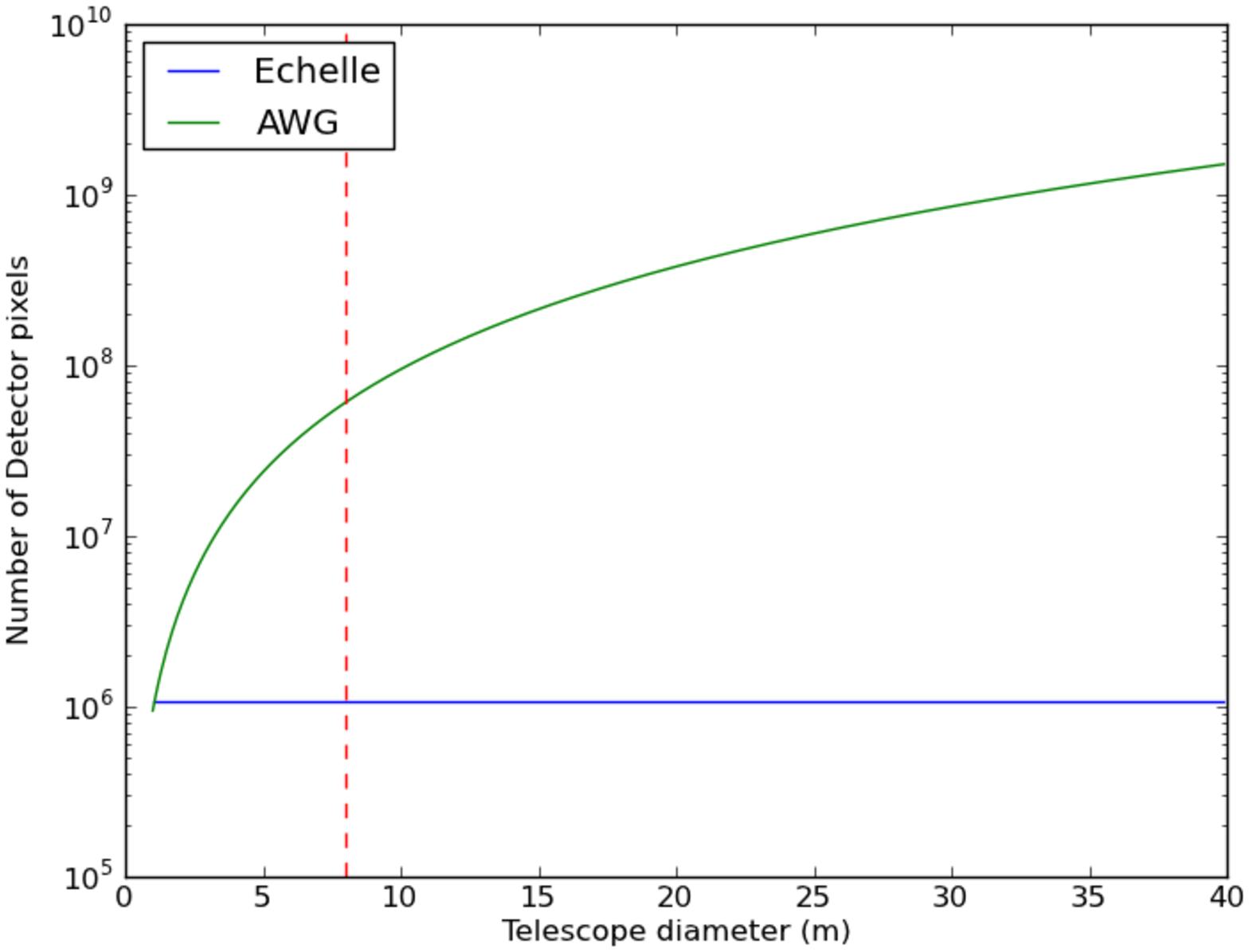}
			\caption{The resulting pixel numbers due to varying the diameter of the telescope feeding our model instrument. The different in number of pixels increases greatly when the telescope diameter increases. The dashed line represents GNIRS.}
			\label{fig:d_tel_pixels}
		\end{center}
	\end{minipage}
	\hfill
\end{figure}

\begin{figure}[h]
	\hfill
	\begin{minipage}[t]{.45\textwidth}
		\begin{center}
			\includegraphics[width = 1.0\textwidth]{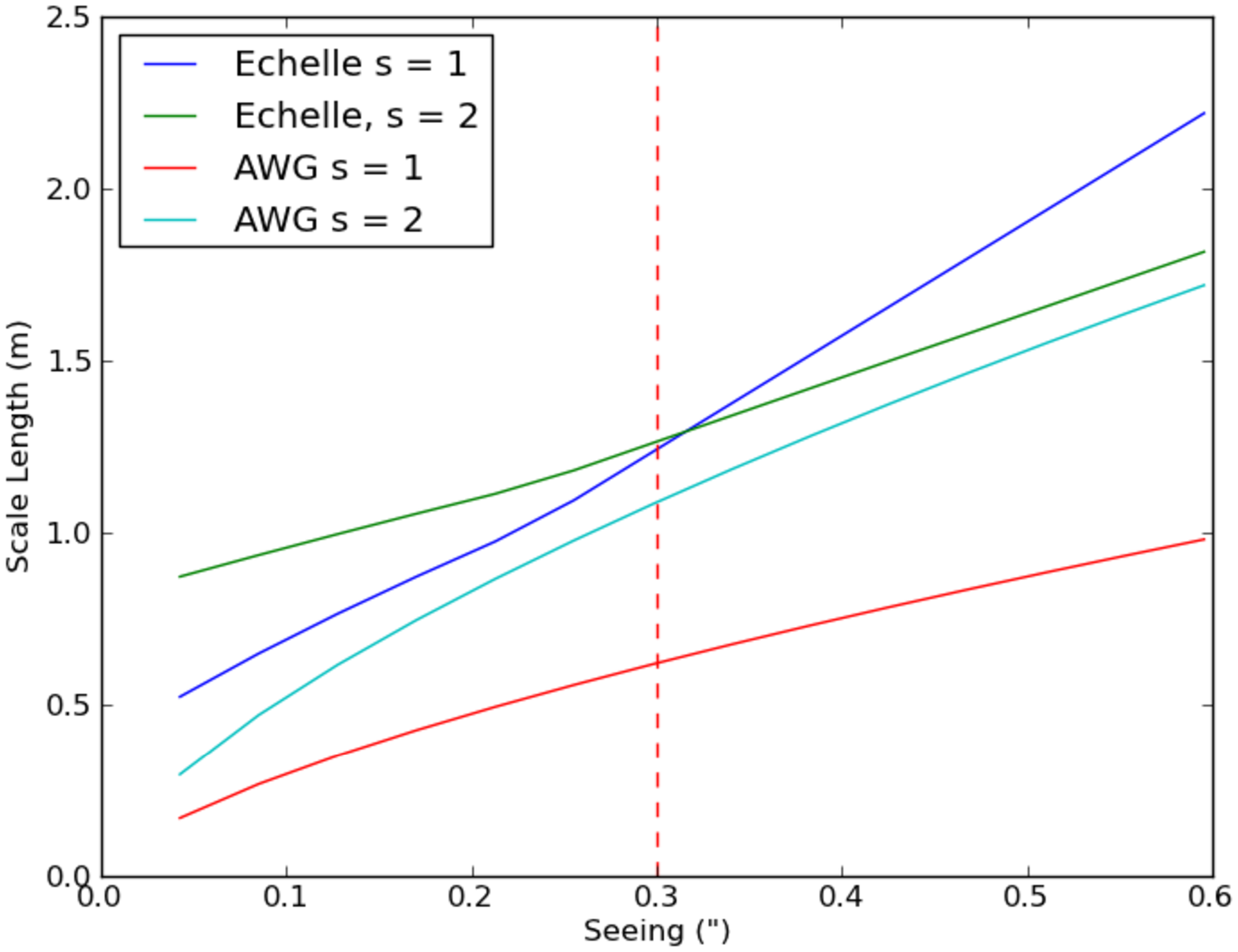}
			\caption{The resulting scale lengths due to varying the seeing of the telescope. The advantage of using an AWG instrument is greatest when the input is close to the diffraction limit. The dashed line represents GNIRS.}
			\label{fig:seeing}
		\end{center}
	\end{minipage}
	\hfill
	\begin{minipage}[t]{.45\textwidth}
		\begin{center}
			\includegraphics[width = 1.0\textwidth]{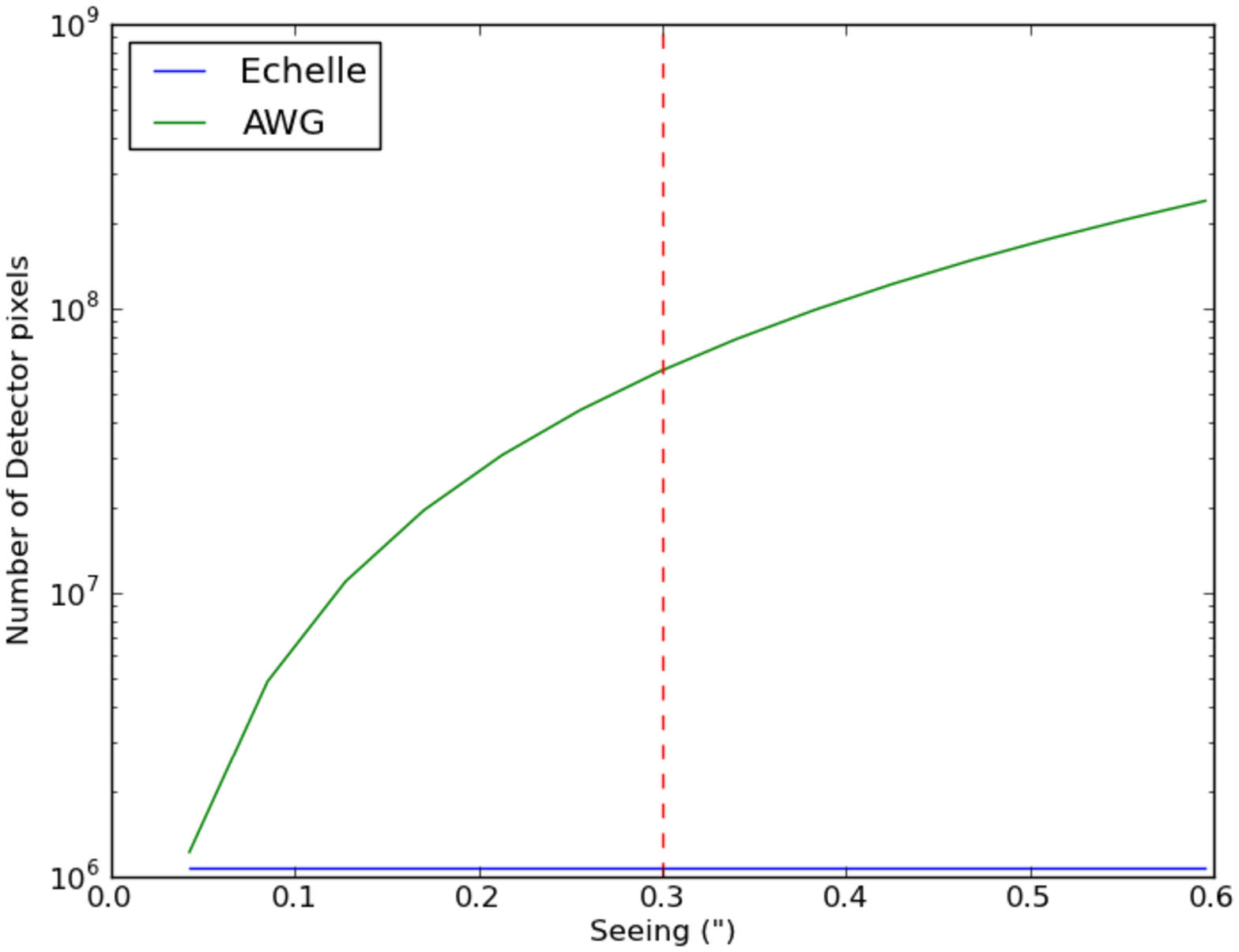}
			\caption{The resulting pixel numbers due to varying the seeing of the telescope. It is easily seen that unless the input is near the diffraction limit the number of pixels in the system will be much greater in the AWG device. The dashed line represents GNIRS.}
			\label{fig:seeing_pixels}
		\end{center}
	\end{minipage}
	\hfill
\end{figure}

The results are presented in graphical form in Figs. \ref{fig:resolution} to \ref{fig:seeing_pixels}. The immediate conclusion is that in all cases the AWG instrument will be smaller. This is slightly misleading though as we have not included detector volumes or instrument housing, which would increase the scale lengths, leading to a much closer comparison. In all cases the number of detector pixels is much higher for the AWG model, though if the input is close to the diffraction limit this number decreases, making the devices more comparable.  \newline

Figure \ref{fig:resolution} shows varying the resolving power will produce similar scale length differences between the AWG and echelle models. This means there is no significant advantage for using the AWG instrument for low or high resolution spectroscopy. In Fig. \ref{fig:res_pixels} we see that using instruments with higher resolving power also results in a similar difference between the AWG and echelle model, suggesting that resolution will not be a major factor in deciding the use of the instrument.  \newline

\begin{figure}
	\begin{center}
		\includegraphics[width = .5\textwidth]{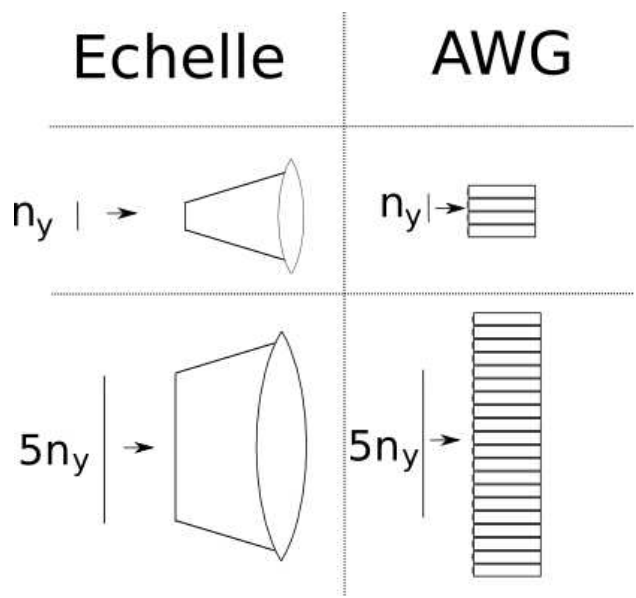}
		\caption{An example of changing the slit length of the different model instruments. using a slit length of $n_{y}$ yields a larger echelle instrument, though increasing the slit length by a factor of 5 makes the AWG instrument larger.}  
		\label{fig:awg_ech_comp}
	\end{center}
\end{figure}

Figure \ref{fig:fov} suggests the size of the AWG instrument will have the greatest advantage for smaller fields of view. The reason for this is that for small fields of view the angular dependence of the echelle makes the overall instrument larger than the AWG. When the slit length is increased the increase in the angle for the echelle becomes smaller than the extra size of the extra AWGs. This is illustrated in Fig. \ref{fig:awg_ech_comp}. Figure \ref{fig:fov_pixels} shows once again the number of detector pixels on the AWG instrument becomes larger as the FOV increases, but this is at the same rate as the echelle instrument. \newline

Figure \ref{fig:d_tel} suggests increasing the size of the telescope will not make any significant difference to the size difference of the AWG and echelle models. This means the same size issues will be present for AWG instruments as conventional ones for ELTs using the model setup. We also see from Fig. \ref{fig:d_tel_pixels} that as we increase the telescope diameter the number of pixels required increases with the number of propagating modes(see Eq. (\ref{eqn:modes})). \newline

Figure \ref{fig:seeing} suggest that the relative sizes of the instruments is not highly dependent on seeing. This should be qualified though. If the input is at the diffraction limit there will be no need for a photonic  lantern, or potentially a smaller one, leading to a smaller and less complicated instrument, if the input is far from the diffraction limit it will require a large photonic lantern. Figure \ref{fig:seeing_pixels} shows the number of detector pixels is comparable for diffraction limited inputs, but increases hugely as the seeing becomes worse, due to the extra modes (see \ref{eqn:modes}). \newline

%

\section{Conclusion}

We have used simplified models for both a conventional echelle model and an AWG model to simulate the size and number of detector pixels of GNIRS on Gemini. We have then varied resolution, telescope diameter, sample size and seeing. Our results are presented in terms of the scale length (cube root of the volume) of the instrument and the number of detector pixels the instrument will require.

We find that the AWG model will be slightly smaller in size for all cases becoming much smaller when the FOV of the instrument is small. This is slightly misleading though as we have not included other components in the model (such as detectors and housing). This means the actual instrument will most likely be the same size if not larger than a conventional non-photonic counterpart. We also find this problem will be compounded for larger telescopes unless the input is very close to the diffraction limit. We also find that larger a larger FOV makes the AWG instrument less competitive in terms of size, meaning that AWG instruments may be suited to instruments with smaller fields of view, such as amateur instruments or for high resolution spectroscopy. \newline
Our results indicate the number of detector pixels will be substantially more for photonic instruments unless the input is very close to the diffraction limit. This suggests that either near diffraction limited Adaptive Optics or a space based mission would be required for optimal performance from photonic spectrographs.

In conclusion, our findings suggest that photonic spectroscopy would be most useful on small instruments with diffraction limited seeing. We find that the AWG instrument will be on the same magnitude in terms of size as a conventional spectrograph, but will have more detector pixels. \newline
This does not provide a comprehensive model as we have only studied photonic spectrographs with a single input. Recent Astrophotonics papers \cite{Bland-Hawthorn2010, Cvetojevic2012}  suggest that multiple inputs could be used with varying configurations. This would potentially reduce the size of the instrument, but would not reduce the number of detector pixels required unless all the spectra could be combined onto a single detector pixel. We intend to investigate this further with similar models in the near future.

\section{Acknowledgements}
R. Harris gratefully acknowledges support from the Science and Technology Facilities council. J.R. Allington-Smith wishes to thank the European Union for funding for the OPTICON Research Infrastructure for Optical/IR  astronomy under Framework Programme 7. We also thank Robert Thomson for many useful discussions into Photonic devices.

\bibliographystyle{spiebib}
\bibliography{ref}

\begin{thebibliography}{10}

\bibitem{Lee2000}
D.~Lee and J.~R. Allington-Smith, ``{An experimental investigation of immersed
  gratings},'' {\em Monthly Notices of the Royal Astronomical Society}~{\bf
  312}, pp.~57--69, Feb. 2000.

\bibitem{Bland-Hawthorn2006}
J.~Bland-Hawthorn and A.~Horton, ``{Instruments without optics: an integrated
  photonic spectrograph},'' {\em Proceedings of SPIE}~{\bf 6269}, pp.~21--34,
  June 2006.

\bibitem{Allington-Smith2009}
J.~Allington-Smith and J.~Bland-Hawthorn, ``{Astrophotonic spectroscopy:
  defining the potential advantage},'' {\em Monthly Notices of the Royal
  Astronomical Society}~{\bf 404}, pp.~232 -- 238, Oct. 2010.

\bibitem{Bland-Hawthorn2010}
J.~{Bland-Hawthorn}, J.~{Lawrence}, G.~{Robertson}, S.~{Campbell}, B.~{Pope},
  C.~{Betters}, S.~{Leon-Saval}, T.~{Birks}, R.~{Haynes}, N.~{Cvetojevic}, and
  N.~{Jovanovic}, ``{PIMMS: photonic integrated multimode microspectrograph},''
  in {\em Society of Photo-Optical Instrumentation Engineers (SPIE) Conference
  Series},  {\em Society of Photo-Optical Instrumentation Engineers (SPIE)
  Conference Series} {\bf 7735}, July 2010.

\bibitem{Noordegraaf2010}
D.~Noordegraaf, P.~M.~W. Skovgaard, M.~D. Maack, J.~Bland-Hawthorn, R.~Haynes,
  and J.~Laegsgaard, ``{Multi-mode to single-mode conversion in a 61 port
  Photonic Lantern.},'' {\em Optics express}~{\bf 18}, pp.~4673--8, Mar. 2010.

\bibitem{Cvetojevic2009}
N.~{Cvetojevic}, J.~S. {Lawrence}, S.~C. {Ellis}, J.~{Bland-Hawthorn},
  R.~{Haynes}, and A.~{Horton}, ``{Characterization and on-sky demonstration of
  an integrated photonic spectrograph for astronomy},'' {\em Optics
  Express}~{\bf 17}, pp.~18643--18650, Oct. 2009.

\bibitem{Cvetojevic2012}
N.~{Cvetojevic}, N.~{Jovanovic}, J.~{Lawrence}, M.~{Withford}, and
  J.~{Bland-Hawthorn}, ``{Developing arrayed waveguide grating spectrographs
  for multi-object astronomical spectroscopy},'' {\em Optics Express}~{\bf 20},
  p.~2062, Jan. 2012.

\bibitem{Lawrence2010a}
J.~{Lawrence}, J.~{Bland-Hawthorn}, N.~{Cvetojevic}, R.~{Haynes}, and
  N.~{Jovanovic}, ``{Miniature astronomical spectrographs using
  arrayed-waveguide gratings: capabilities and limitations},'' in {\em Society
  of Photo-Optical Instrumentation Engineers (SPIE) Conference Series},  {\em
  Society of Photo-Optical Instrumentation Engineers (SPIE) Conference Series}
  {\bf 7739}, July 2010.

\bibitem{cheo}
P.~K. Cheo, {\em Fiber optics and Optoelectronics}, Prentice-Hall, Englewood
  Cliffs, New Jersey, 1990.

\bibitem{Peralta2003}
L.~G.~D. Peralta, A.~A. Bernussi, S.~Frisbie, R.~Gale, and H.~Temkin,
  ``{Reflective Arrayed Waveguide Grating Multiplexer},'' {\em IEEE}~{\bf
  15}(10), pp.~1398--1400, 2003.

\bibitem{DePeralta2004}
L.~de~Peralta, a.a. Bernussi, V.~Gorbounov, J.~Berg, and H.~Temkin, ``{Control
  of center wavelength in reflective-arrayed waveguide-grating multiplexers},''
  {\em IEEE Journal of Quantum Electronics}~{\bf 40}, pp.~1725--1731, Dec.
  2004.

\bibitem{Allington-Smith2009a}
J.~R. {Allington-Smith}, ``{Strategies for spectroscopy on Extremely Large
  Telescopes - I. Image slicing},'' {\em Monthly Notices of the Royal
  Astronomical Society}~{\bf 376}, pp.~1099--1108, Apr. 2007.

\bibitem{2006MNRAS.371..380A}
J.~R. {Allington-Smith}, R.~{Content}, C.~M. {Dubbeldam}, D.~J. {Robertson},
  and W.~{Preuss}, ``{New techniques for integral field spectroscopy - I.
  Design, construction and testing of the GNIRS IFU},'' {\em MNRAS}~{\bf 371},
  pp.~380--394, Sept. 2006.

\end{thebibliography}

\begin{table}
	\caption{Table of input parameters for GNIRS calibration, taken from Ref. \citenum{Allington-Smith2009a} and calculated using the model. These will be used as the default parameters for the models.}
	\label{tab:instruments}
	\begin{tabular}{ | c | c | c | c | c | c | c | c | c | c | c |}  \hline 
	 	 $\chi$ & Slit & Telescope   & R & $\lambda_{c}$  & $\Delta_{FSR}$  & Vol &     $F_{tel}$/$F_{cam}$ & S & a & b \\  
 		  (") &Length (")  & Diameter (m)   &   & (nm) &  (nm) & $(m^{3})$  & (m) & & (mm) &  \\ \hline
		  \mr{0.3} & \mr{99} & \mr{8} & \mr{5900} & \mr{1650} & \mr{400} & \mr{2.00} &  \mr{16} & 1  & 100 & 1.95 \\  \cline{9-11}
		  & & & & & & &  & 2 & 500 & 1.1 \\ \hline		
	\end{tabular} 	

\end{table}

\end{document}